\documentstyle[sprocl]{article}

\input{psfig}
\def\be{\begin{equation}}
\def\ee{\end{equation}}
\def\bea{\begin{eqnarray}}
\def\eea{\end{eqnarray}}

\newcommand{\dslash}{\partial\hspace{-.09in}/}

\newcommand{\pslash}{p\hspace{-.07in}/}
\newcommand{\Dslash}{D\hspace{-.11in}/}

\begin{document}

\title{Coherent Neutrino Propagation in a Dense Medium}

\author{Ken Kiers}
\address{Department of Physics,Brookhaven National Laboratory}
\author{Nathan Weiss\footnote{Talk presented at ``Strong \& Electroweak
Matter'' in Eger, Hungary; May 1977}}
\address{ Tel Aviv University, Weizmann Institute and
University of B.C.}

\maketitle

\begin{abstract}
Motivated by the effect of matter on neutrino oscillations (the MSW
effect)
we have studied the propagation of neutrinos in a dense medium.
The dispersion relation for massive neutrinos in a medium is known
to have a minimum at nonzero momentum $p\sim G_F\rho/\sqrt{2}$.
We have studied in detail the origin and consequences of this dispersion
relation
for both Dirac and Majorana neutrinos both in a toy model with only
neutral currents and a single neutrino flavour and in a realistic 
``Standard Model'' with two neutrino flavours. For a range
of neutrino momenta near the minimum of the dispersion relation, Dirac
neutrinos are trapped by their coherent interactions with the medium.
This effect does not lead to the trapping of Majorana neutrinos.
\end{abstract}

 Motivated by the effect of matter on neutrino oscillations (the 
MSW effect~\cite{msw}),
there have been several works in recent years aimed at understanding
in a more complete way the propagation of one or more flavours of
massive
neutrinos in matter (such as in the Sun, in Neutron Stars or in 
Supernovae for which the density difference 
$\langle\rho_{e^-}-\rho_{e^+}\rangle\ne 0$). 
In our recent paper \cite{knw}we presented the results of our study
of some of the unusual features of this problem. We were
 particularly
interested in effects at low neutrino momentum and in effects
due to the minimum which
the dispersion relation has at nonzero momentum. 
In this presentation we begin   by examining in some detail a 
simple model with only a single neutrino flavour
in which the neutrino propagates in a background of
electrons.  This model will show many of the essential features
of the more realistic models. We shall then simply state the
results in these other cases including the case of the Standard
Model both with Dirac and Majorana neutrinos. The reader is 
referred to our paper \cite{knw} for more details.

We begin by considering a simplified
model in which a Dirac neutrino propagates in an electron ``gas'' to
which it couples only via the neutral current interaction. 
Our model may be described by the following
Lagrangian 
\begin{eqnarray}
        {\cal L} ={\overline \psi}_{\nu}
		\left( i\Dslash^+ - m_{\nu}\right)\psi_{\nu}
		+{\overline \psi}_e
                \left( i\Dslash^- - m_e\right)\psi_e
		 -\frac{1}{4}F^2+\frac{m_Z^2}{2}Z^2
		-\mu_e\psi_e^{\dagger}\psi_e ,
	\label{eq:ltoy}
\end{eqnarray}
where
\begin{eqnarray}
	D_{\mu}^{\pm} & = & \partial_{\mu} \pm i\frac{g}{2\sqrt{2}}
		Z_{\mu}(1-\gamma^5), \\
	F_{\mu\nu} & = & \partial_{\mu}Z_{\nu}-\partial_{\nu}Z_{\mu}.
\end{eqnarray}
The chemical potential term in the Lagrangian is included 
in order to give a non-zero value to the electron density; that is
\equation
	\rho_e = \langle \psi_e^{\dagger}\psi_e \rangle \ne 0.
\endequation

A detailed analysis of the above theory is given in our paper\cite{knw}.
An interesting way to gain insight into the Physics of our model
is by examining the equations of motion following from
the Lagrangian in Eq.~(\ref{eq:ltoy}).  Varying the Lagrangian
with respect to the $Z^{\mu}$ field leads to
\equation
	\partial_{\nu}F^{\nu\mu} + m_Z^2Z^{\mu} = - \frac{g}{2\sqrt{2}} 
			J^{\mu},
	\label{zmotion}
\endequation
where
\equation
	J^{\mu} =\overline{\psi}_e\gamma^{\mu}(1-\gamma^5)\psi_e 
		-\overline{\psi}_{\nu}\gamma^{\mu}(1-\gamma^5)\psi_{\nu}.
\endequation
If $\rho_e$$=$$\langle\psi_e^{\dagger}\psi_e\rangle$ is constant,
then (\ref{zmotion}) leads to
\equation
	\langle Z^0\rangle = - \frac{g\rho_e}{2\sqrt{2}m_Z^2},
	\label{zexp}
\endequation
that is, the $Z^0$ field has gained a vacuum expectation
value.  The equation of motion for the neutrino field
is then given by
\equation
	\left[ i\dslash -m +\alpha\gamma^0(1-\gamma^5)
		\right]\psi_{\nu} = 0,
\endequation
where $\alpha=g^2\rho_e/8m_Z^2$. 
From this point of view the left-handed neutrino
sees a 
mean (coherent) ``scalar potential,'' $\langle Z^0\rangle$. 

For constant electron density, the presence of the ``chiral potential''
in this expression leads to a {\em shift} in the frequency
by $\alpha$, but only for the left-handed (chiral) piece.  This 
shift in the frequency is precisely the 
``index of refraction'' familiar from the MSW effect.
The shift in energy for the neutrino is opposite that for the
anti-neutrino
\footnote{This shift comes from a term in the effective Hamiltonian 
proportional to $\psi^\dagger_\nu\psi_\nu$ which equals the number
density
of neutrinos minus the number density of antineutrinos.}.
If the neutrino is ``attracted'' by the medium, then the anti-neutrino
is ``repelled'' by it.

The above Dirac Equation leads to four solutions for the energy:
\equation
	\omega = -\alpha \pm \sqrt{(p+\alpha s)^2+m^2}.
	\label{energies}
\endequation
These dispersion relations are plotted in Figure 1
both for $m$$=$$0$ (dashed curves)  and $m$$\ne$$0$
(solid curves.)  Several key features of these plots
should be noted.  First of all, the ``negative energy'' states
are, in this case, those which are unbounded from below as 
the momentum is increased.  In the second quantized theory the
correct energy of such a state is just the negative of
its energy eigenvalue.  We also note that when 
$m$$=$$0$ there are ``level crossings.'' 
These are avoided for $m$$\ne$$0$ by level repulsion
due to the mixing of the levels.

\centerline{\psfig{figure=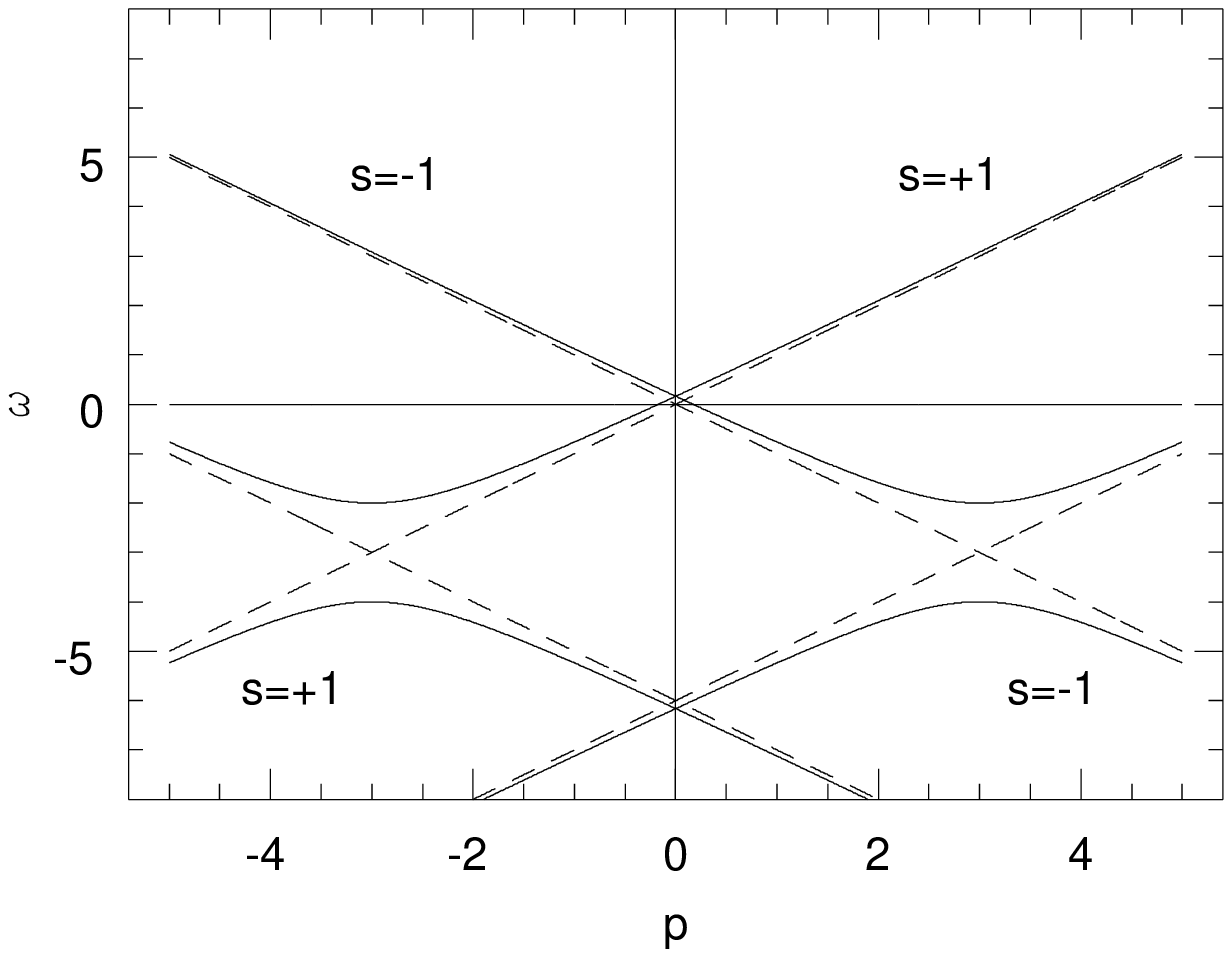,height=4in}}
\vskip -1.9in
\centerline{\it Figure 1: Dispersion relations for a neutrino in an 
electron ``gas''}

\medskip

Note that
the  minima of some of the dispersion relations occur at
non-zero values of the momentum, $p$$=$$\pm\alpha$, instead
of at the origin.
In fact the minimum energy $\omega_{\rm min}$$=$$-\alpha+m$ is
{\em less than} the neutrino mass.  Thus it is possible to
produce a neutrino in the medium which has 
$\omega$$<$$m$.  Such a neutrino will not
have enough energy to survive in the vacuum and will thus be
{\em trapped} by the medium. 
The condition for trapping is\cite{knw}:
\equation
	p < p_{\rm trap} \equiv \alpha + \sqrt{\alpha^2 + 2\alpha m}.
	\label{ptrapdef}
\endequation
Thus, neutrinos produced in this medium with momentum $p$$<$$p_{\rm
trap}$
will not have enough energy to survive in the vacuum and will
be trapped.   

Before proceeding it is useful to get some idea
of the overall magnitude of the effect of neutrino trapping.
Setting $G$$\approx$$G_F$ and $m$$\approx$$10^{-3}$eV (which is 
a mass relevant for the MSW-resolution of the solar neutrino problem),
we find that
$p_{\rm trap}$$\sim$$10^{-8}$eV in the sun (for
which $\alpha$$\sim$$10^{-12}$eV) and 
$p_{\rm trap}$$\sim$$100$eV in a supernova (for
which $\alpha$$\sim$$100$eV.) 
 
In the case of Majorana neutrinos, there is only a single
(left-handed) field, $\chi_L$.
The dispersion relations in this case 
can be obtained by solving the equations of motion, as was
first done by Mannheim~\cite{mannheim}.  
The   dispersion relation for the negative helicity neutrino is given by
\equation
	\omega = \pm\sqrt{\left(|\vec{p}|-2\alpha\right)^2+m^2}.
\endequation
In this case the energy has a minimum value,
$\omega$$=$$m$, which occurs at $|\vec{p}|$$=$$2\alpha$.
In fact, the dispersion relation in matter is identical to that in
vacuum except for a lateral shift to the right.  This implies in
particular that, in contradistinction to the Dirac case,
a neutrino cannot have an energy less than $m$ and 
there is thus {\em no trapping} of Majorana neutrinos
in the medium.   

We turn now to  a more realistic case in which 
there are two neutrino flavours and in which there are both
neutral current and charged current couplings to the medium.
We have in mind, of course, the Standard Electroweak Model
with massive neutrinos.
The Dirac Equation in this case is given by:
\equation
	\left\{\pslash -M +\left(\beta-\alpha Q\right) \gamma^0\left(
	1-\gamma_5\right)\right\}\psi=0
	\label{deqn2neut}
\endequation
where $\psi$ has two flavour components,
$M$ is the {\bf diagonal} 
$2\times 2$ mass matrix, $\beta\propto \rho G_F$
is the contribution of the neutral current which couples only to the
left
handed neutrinos, $\alpha\propto \rho_e G_F$ represents the charged
current
contribution which couples only to $\nu_e$ and $Q$ is
the 
mixing matrix
\equation
	Q=\left(\matrix{\cos^2(\theta)& \sin(\theta)\cos(\theta)\cr
	\sin(\theta)\cos(\theta)&\sin^2(\theta)}\right) .
	\label{qdef}
\endequation

This Dirac equation leads to the following quartic equation
for the dispersion relation:
\equation
	\left[\omega^2-p^2-\mu^2+(2\beta-\alpha)(\omega-sp)\right]^2
	=\alpha^2(\omega-sp)^2-\alpha\Delta^2\cos(2\theta)(\omega-sp)
	+\frac{1}{4}\Delta^4 ,
	\label{quartic}
\endequation
in which $s$$=$$\pm 1$ is the eigenvalue of $\sigma_3$, the spin
projection in the $+z$ direction.  The eight
solutions to this equation
(for $s$$=$$\pm 1$) lead to the eight dispersion relations (four
positive
energy and four negative energy) in the medium.

It is easy to see that both in the massless case and in the
case when $\theta=0$ the solutions to the quartic equation
are precisely the dispersion relations expected in these cases.
In order to analyze
the dispersion relations when the
coupling $\theta$ is nonzero but not too large it is to use a graphical
approach.  One begins by looking at the solutions when $\theta$$=$$0$
in which case the two neutrino flavours decouple.
Thus, for example, in Figure 2(a) the dotted curves
represent the solutions for $\theta$$=$$0$. Note that the dispersion
relations for the two flavours of neutrinos cross at some points.
When $\theta$ is ``turned on'' we expect that these levels will repel
and will lead to a curve similar to the solid curve in that figure.
The solid curve is, in fact, the solution to the quartic equation when
$\theta$$=$$0.2$. This graphical method is reasonably accurate when
$\theta$
is small but can be used as a guide even for larger values of $\theta$.
Notice that just as for the single neutrino case, the lightest
Dirac neutrino is trapped by the medium.

\centerline{\psfig{figure=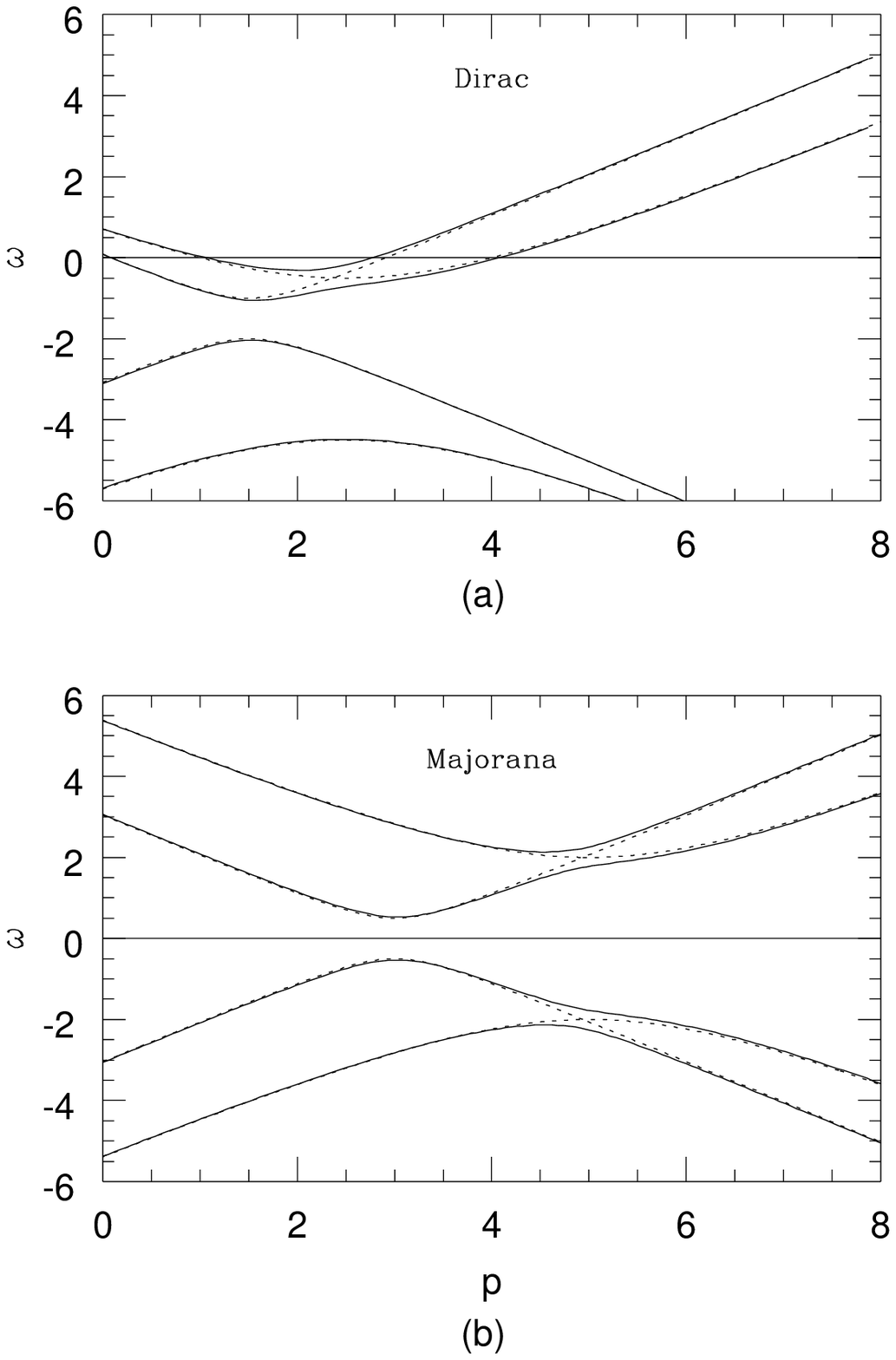,height=3.6in}}
\vskip -.4in
\centerline{\it Figure 2: Dispersion relations for two neutrino flavours.}

\medskip

The case of Majorana neutrinos is interesting for two 
reasons.  First of all, it is the favoured realistic scenario
in models which have massive neutrinos, for example in models
which employ the ``see-saw'' mechanism.  Secondly 
the equations governing the dispersion relations 
are {\em quadratic} rather than {\em quartic}. See \cite{knw} for details.
Analysis of these dispersion relations shows that
their minimum
is always greater than or equal to the minimum mass
and so again there appears to be no trapping in the Majorana case.

\vglue -.2in


\section*{References}
\begin{thebibliography}{99}

\bibitem{msw} L. Wolfenstein, Phys. Rev. D {\bf 17} (1978) 2369;
	Phys. Rev. D {\bf 20} (1979) 2634;
	S.P. Mikheyev and A.Yu. Smirnov, Yad. Fiz. {\bf 42} (1985)
	1441 [Sov. J. Nucl. Phys. {\bf 42} (1985) 913]; Il Nuovo 
	Cimento C {\bf 9} (1986) 17.
\bibitem{knw} K. Kiers and N. Weiss, hep-ph 9704346; 
   to appear in Phys. Rev. D 
\bibitem{mannheim} P.D. Mannheim, Phys. Rev. D{\bf 37} (1988) 1935.
\end{thebibliography}
\end{document}